\documentclass[aps,pra,reprint,amsmath,amssymb,floatfix,citeautoscript,superscriptaddress]{revtex4-1}
\usepackage{amsmath}
\usepackage{graphicx}
\usepackage[usenames,dvipsnames]{xcolor}
\definecolor{myblue}{rgb}{0,0,1}
\definecolor{mywhite}{rgb}{1,1,1}
\usepackage[colorlinks=true,linkcolor=myblue,citecolor=myblue]{hyperref}

\usepackage[T1]{fontenc}
\usepackage{txfonts}

\def\t{\textrm}
\begin{document}
\title{Description of quasiparticle and satellite properties via 
cumulant expansions of the retarded one-particle Green's 
function}

\author{Matthew Z. Mayers}
\affiliation{Department of Chemistry, Columbia University, New York, NY 10027, USA}
\author{Mark S. Hybertsen}
\affiliation{Center for Functional Nanomaterials, Brookhaven National Laboratory, Upton, NY 11973-5000, USA}
\author{David R. Reichman}
\affiliation{Department of Chemistry, Columbia University, New York, NY 10027, USA}

\begin{abstract}
A new cumulant-based $GW$ approximation for the retarded one-particle 
Green's function is proposed, motivated by an exact relation between the 
improper Dyson self-energy and the cumulant generating function. 
Qualitative aspects of this method are explored within a simple one-electron 
independent phonon model, where it is seen that the method preserves 
the energy moment of the spectral weight while also reproducing the exact 
Green's function in the weak coupling limit. For the three-dimensional 
electron gas, this method predicts multiple satellites at the bottom of the 
band, albeit with inaccurate peak spacing. However, its quasiparticle 
properties and correlation energies are more accurate than both previous 
cumulant methods and standard $G_0W_0$. Our results point to new 
features that may be exploited within the framework of cumulant-based 
methods and suggest promising directions for future exploration and 
improvement of cumulant-based $GW$ approaches.
\end{abstract}

\maketitle

\section{Introduction}

The development of methods that can accurately and affordably describe 
both the total electronic energy and the electronic excitations of complex 
systems remains a long-standing challenge in both condensed matter 
physics and chemistry. Substantial progress has been made 
in recent decades on the development of approximate approaches 
to calculate correlation contributions that go beyond the Hartree-Fock 
level of mean-field theory. While density functional theory (DFT), including 
its extensions to hybrid functionals, has emerged as an accurate and efficient means 
of calculating many properties of both solids and molecules, systematic improvement 
of DFT is challenging~\cite{Cohen12}. In particular, while the Kohn-Sham eigenvalues 
in the theory often give a surprisingly useful band structure, there are fundamental 
differences with respect to properly calculated excitation energies~\cite{Onida02,Cohen12}.
More broadly, considering the proliferation of a myriad of approximate 
exchange-correlation functionals, care must be taken in applications to 
assess empirical evidence of the accuracy for specific classes of materials. 

Separate from DFT, direct many-body methods based on wavefunctions 
have achieved impressive accuracy, exemplified by coupled-cluster 
methods for finite systems~\cite{Bartlett07} and quantum Monte Carlo 
(QMC) methods for extended systems~\cite{Foulkes01}. These approaches are very 
challenging numerically due to unfavorable scaling with system size (or complexity), 
but are often regarded as a "gold standard" when they can be applied. They are 
also typically more difficult to apply to excited state properties with the same accuracy.
Nonetheless, recent progress is encouraging for more widespread application to solids~\cite{Booth13,Shepherd13,McClain15}.

On the other hand, Green's function-based perturbation expansions 
by their very nature describe spectral features and quasiparticle properties 
of extended systems in a size-consistent manner~\cite{Hedin1969}. 
In particular, the development of the $GW$ approximation for application
to the one-particle Green's function~\cite{Hedin65} led to the first predictive 
calculations of charged excitation energies in real materials
~\cite{Hybertsen86,Godby87,Ary:1,Aulbur1999}. The extension to the 
Bethe-Salpeter equation for the two-particle Green's function
has correspondingly supported calculation of the neutral excitations,
such as those probed in optical absorption~\cite{Onida02}. The development 
of systematic corrections beyond $GW$, including approximations to the vertex 
corrections and the use of self-consistency, remains a subject of ongoing research~\cite{Shirley96,Holm96,Holm98,Bruneval05,Schilfgaarde06,Shishkin07,Guzzo11,Gruneis14}.
Interestingly, the corresponding Green's function based method 
for the total electronic energy has not been widely used,
although several formulations have been investigated~\cite{Holm98,Holm00,Dahlen06}.

The homogeneous electron gas model in three dimensions (3D), 
capturing essential features of the electronic structure of simple metals,
has been widely used as a model system. Results based on $G_0W_0$ 
(the non-self-consistent first iteration of the $GW$ approximation) show 
very reasonable quasiparticle properties, but a satellite structure 
(``plasmaron peak'') about $1.5\omega_p$ below the quasiparticle 
peaks ($\omega_p$ being the plasma energy)~\cite{Lundqvist68}. This 
is a surprising result since standard coupled electron-boson models would 
suggest a series of satellite peaks near integer multiples of $\omega_p$~\cite{Mahan}
below the quasiparticle peaks. Calculations in which $G$ was iterated to 
self-consistency, conceptually part of Hedin's original framework~\cite{Hedin65},
indicated further unphysical changes in the satellite region~\cite{Shirley96,Holm96}.
Interestingly, self-consistent $GW$ gave reasonable correlation energies,
but it was suggested that vertex corrections were needed in addition to restore
physical spectral properties~\cite{Holm00}.

The difficulty of describing satellite structures in the spectral function
strongly suggests a cumulant-based approach
to approximately include vertex corrections~\cite{Kubo:8}.
This idea has been extensively explored with a time-ordered formulation of $G$
~\cite{Hedin80,Almbladh83,Aryasetiawan96,Holm97,Vos02,Guzzo11,Lischner14},
and has been successful in describing the satellite structure in 
metals and semiconductors~\cite{Lischner13,Guzzo14,Caruso15a, Caruso15b,Zhou15}.
The approach restores the expected satellite progression
and modifies the quasiparticle properties quantitatively.
In part, the exponential form imposed by the cumulant ansatz 
leads to the inclusion of higher-order exchange-like diagrams that 
are only accessible in the standard $GW$ formalism by way of 
vertex corrections. However, these higher-order diagrams do \textit{not} 
correspond exactly to standard diagrams in the time-ordered Dyson 
expansion. More generally, the cumulant approach has not yet reached 
the formal level of sophistication that is afforded by the standard 
diagrammatic apparatus that surrounds the Dyson equation. In particular, 
aspects related to self-consistency, conservation laws, and the one-to-one 
correspondence of terms within the cumulant expansion to standard 
Feynman-Dyson diagrammatics require further investigation.

The time-ordered cumulant approach is limited by the serious drawback 
that it precludes the possibility of positive spectral weight both above and 
below the chemical potential. Recently, Kas \emph{et al.} showed that 
the retarded Green's function is a more natural quantity to employ with 
the cumulant formalism, as it allows for the description of both particles 
and holes within one spectral weight profile~\cite{Kas14}. Compared with standard 
$G_0W_0$~\cite{Lundqvist68}, the retarded cumulant approach predicts 
more physical satellite properties, similar energies, but somewhat less 
accurate wavevector-dependent occupation numbers for the 3D 
electron gas. In this work, we take this retarded Green's function perspective 
as a starting point to investigate and compare new cumulant-based $GW$ 
schemes.

\section{Methodology}

The many-body perturbation expansion for the one-particle Green's function can be resummed via the Dyson equation,
\begin{subequations}
\label{dyson}
\begin{eqnarray}
\nonumber G_k(\omega) &=& G_k^0(\omega) + G_k^0(\omega)\Sigma_k^I(\omega)G_k^0(\omega),\\
&=& G_k^0(\omega)[1 + \Sigma_k^{I}(\omega)G_k^0(\omega)] \label{dysoni}\\
&=& {G_k^0(\omega) \over 1 - \Sigma_k^*(\omega)G_k^0(\omega)} \label{dysonp},
\end{eqnarray}
\end{subequations}

\noindent where $G_k^0(\omega)$ is the non-interacting Green's function, $\Sigma_k^*(\omega)$ is the proper self-energy, and $\Sigma_k^I(\omega)$ is the improper self-energy~\cite{Hedin1969}. In the non-self-consistent $GW$ approximation (henceforth referred to as $G_0W_0$), $\Sigma_k^*(\omega)$ is truncated at first order, and the random phase approximation (RPA) $W_k(\omega)$ is used in place of the bare Coulomb interaction $v_k$~\cite{Hedin1969}:

\begin{subequations}
\begin{eqnarray}
\Sigma_k^*(\omega) &=& {i \over \hbar}{1 \over (2\pi)^4}\int d^3qd\omega'G_k(\omega)W_{k - q}(\omega - \omega')\\
W_k(\omega) &=& {v_k \over 1 - \Pi_k(\omega)v_k}\\
\Pi_k(\omega) &=& {i \over \hbar}{1 \over (2\pi)^4}\int d^3qd\omega' G_q(\omega)G_{k + q}(\omega - \omega').
\end{eqnarray} 
\end{subequations}

The retarded cumulant ansatz is a resummation of Eqs.~(\ref{dyson}). It can be written as~\cite{Kas14}
\begin{equation}
G_k^R(t,T) = G_k^{0,R}(t,T)e^{C_k^R(t,T)},
\label{ansatz}
\end{equation}

\noindent where $C_k(t,T)$ is the time-local cumulant function and the `$R$' superscripts denote retarded quantities. 

When considered with Eq. \ref{dysoni}, the cumulant ansatz for the retarded one-particle 
Green's function leads to a simple closed and exact relation between 
the \emph{improper} Dyson self-energy and the cumulant generating function:

\begin{eqnarray}
C_k(t,t') &=& \ln\left(1 + \left[G_k^{R,0}(t,t')\right]^{-1}\iint dt_1dt_2 \times\right.\nonumber\\
&&\hspace{40pt}\left.\textcolor{mywhite}{\int}G_k^{R,0}(t,t_1)\Sigma_k^{R,I}(t_1,t_2)G_k^{R,0}(t_2,t')\right).
\label{cum_time}
\end{eqnarray}

\noindent For simplicity, the expressions are written for the electron gas model.
While it is clear that Eq.~\ref{cum_time} trivially reduces to the standard Dyson 
equation, it should be noted that such a simple direct and exact relationship between 
the retarded cumulant and improper retarded Dyson self-energy has, to the best of our knowledge, not 
been noted before. Such a relation is only possible when retarded 
quantities and the improper as opposed to the proper self-energy are used. 
This relation implies new cumulant-like approximations 
distinct from earlier formulations. For example, the lowest order expansion 
of the logarithm in conjunction with  a retarded, improper self-energy 
calculated using the normal first-order $GW$ diagrams yields

\begin{eqnarray}
C_k(t,t') &=& \left[G_k^{R,0}(t,t')\right]^{-1}\times\nonumber\\
&&\hspace{10pt}\iint dt_1dt_2G_{k}^{R,0}(t,t_1)\Sigma_{GW,k}^{R,I}(t_1,t_2)G_k^{R,0}(t_2,t').
\label{cum_irc}
\end{eqnarray}

\noindent This equation for the cumulant (henceforth referred 
to as $G_0W_0$ with improper retarded cumulant, or $G_0W_0$ IRC) 
is superficially nearly identical to the cumulant approach of Kas \emph{et al.} 
($G_0W_0$ with proper retarded cumulant, or $G_0W_0$ PRC)~\cite{Kas14}, 
except that the improper self-energy is used in place of the proper self-energy. 
To calculate the Green's function within one of these two cumulant schemes, 
then, a proper or improper retarded self-energy is first computed as in the 
$G_0W_0$ scheme. Then the self-energy is inserted into the following Fourier-transformed 
version of Eq.~(\ref{cum_irc}) to find the cumulant:
\begin{eqnarray}
C_k(t) &\equiv& C_k(t_0, t_0+t)\nonumber\\
&=& \int d\omega {1\over \pi}{|\t{Im}\Sigma_k^R(\omega + \epsilon_k)| \over \omega^2}(e^{-i\omega t} + i\omega t - 1).
\end{eqnarray}

\noindent Finally, the spectral weight for the Green's function is 
obtained by taking a Fourier transform of Eq.~(\ref{ansatz})~\cite{Kas14}.

Unlike standard $G_0W_0$ and $G_0W_0$ PRC, the cumulant approach 
outlined above no longer sums diagrams in order of the number of interactions, 
and is thus not perturbative in the interaction coupling. Instead, the first 
cumulant in Eq.~\ref{cum_irc} contains diagrams of all orders of the interaction. 
We emphasize that this fact renders the approach neither more or less accurate 
than the more standard $G_0W_0$ and $G_0W_0$ PRC approximations. 
Regardless, the simplicity of the cumulant formalism as outlined above does 
lead to important properties such as positive and normalized spectral 
weight~\cite{Kas14}.

\section{Results}

To gain a first understanding of the implications of the $G_0W_0$ IRC 
approximation, we apply it to the study of a system of independent phonons 
coupled to a single electronic state:
\begin{equation}
H = \sum_k\omega_kb_k^\dagger b_k + c^\dagger c\left[\epsilon_c + \sum_k\lambda_k(b_k^\dagger + b_k)\right],
\label{phononh}
\end{equation}

\noindent where $b_k, b_k^\dagger$ are the annihilation and creation 
operators for the phonon states, $c, c^\dagger$ are those of the electronic 
state, and $\epsilon_c, \lambda_k$ are the excited state electronic energy 
and the phonon coupling, respectively. We utilize an Einstein spectral 
density, $J(\omega) = g\delta(\omega-\omega_c)$.
This is crudely reminiscent of the plasmon spectral density in the 3D 
electron gas. This model has been used previously previously 
in the analysis of interaction effects for core-holes~\cite{Langreth70}
and to develop models for valence band spectral functions~\cite{Holm00}.



For this model Hamiltonian in which the interaction 
propagator $W$ has been replaced by its phonon propagator analogue $D$,
the $G_0D_0$ PRC approach gives the exact result for the spectral weight~\cite{Mahan}. In this 
regard our goal is not to compare with the standard cumulant approach, 
which will of course in this model yield ``better'' results, but to gain an 
intuition for the expected spectral features produced by the IRC method 
as well as to see which features of the approach are likely to be well 
described. 

The dynamical part of the proper Dyson self-energy within this model is
\begin{subequations}
\begin{eqnarray}
\Sigma^*(\omega) &=& {i \over 2\pi}\int_{-\infty}^\infty d\omega' g\omega_c^2G^0(\omega - \omega';\epsilon_c)D^0(\omega';\omega_c)\nonumber\\
&=& {i \over 2\pi}\int_{-\infty}^\infty d\omega' g\omega_c^2{1 \over \omega - \omega' - \epsilon_c + i\delta}\times\nonumber\\
&&\hspace{20pt}\left({1 \over \omega' - \omega_c + i\delta} - {1 \over \omega' + \omega_c - i\delta}\right)\nonumber\\
&=& {g\omega_c^2 \over \omega - \omega_c - \epsilon_c + i\delta},\\
|\t{Im }\Sigma^*(\omega)| &=& \pi g\omega_c^2\delta(\omega - (\omega_c + \epsilon_c)).
\end{eqnarray}
\end{subequations}

\noindent The frequency integration was done by closing the contour in the lower 
imaginary half-plane. The exact and approximate spectral functions are 
then evaluated as outlined in the previous section; the improper self-energy
is described by
\begin{equation}
|\t{Im }\Sigma^I(\omega)| = \pi g(1 + g)\omega_c^2\delta(\omega - \omega_c(1 + g) - \epsilon_c),
\end{equation}

\noindent and the exact and approximate results are
\begin{eqnarray}
A_{\t{PRC}}(\omega) &=& e^{-g}\sum_{l = 0}^\infty{g^l \over l!}\delta(\omega - \epsilon_c + g\omega_0 - \omega_0l),\\
A_{G_0D_0}(\omega) &=& {1\over 1+g}\delta(\omega - \epsilon_c + g\omega_0) + {g \over 1+g}\delta(\omega - \epsilon_c - \omega_0),\hspace{15pt}\\
A_{\t{IRC}}(\omega) &=& e^{-g/(1+g)}\sum_{l=0}^\infty{1 \over l!}\left({g \over 1+g}\right)^l\times\nonumber\\
&& \hspace{63pt}\delta(\omega - \epsilon_c + g\omega_0 - \omega_0(1+g)l),
\end{eqnarray}

\noindent respectively. The exact solution describes a sequence of 
peaks separated by multiples of $\omega_0$ in energy. 
The basic $G_0D_0$ (Dyson) approximation predicts just two peaks, a 
quasiparticle peak and a satellite peak 
separated by $\omega_0(1+g)$, while the $G_0D_0$ IRC formulation 
predicts an infinite series of peaks separated by $\omega_0(1+g)$. 
The $G_0D_0$ IRC spectrum inherits the unphysical spacing from the 
$G_0D_0$ improper self-energy, which becomes correct only in the weak 
coupling ($g \ll 1$) limit. It is easy to check that all three methods give 
normalized spectral weights with an identical first moment
\begin{eqnarray}
\nonumber \left\langle \left|cc^\dagger\left(\epsilon_c + \sum_k\lambda_k(b_k^\dagger + b_k)\right)\right|\right\rangle &=& \int \omega A(\omega)d\omega\\
&=& \epsilon_c,\
\end{eqnarray}

\begin{figure}[t!]
\centering
\includegraphics[scale = 1]{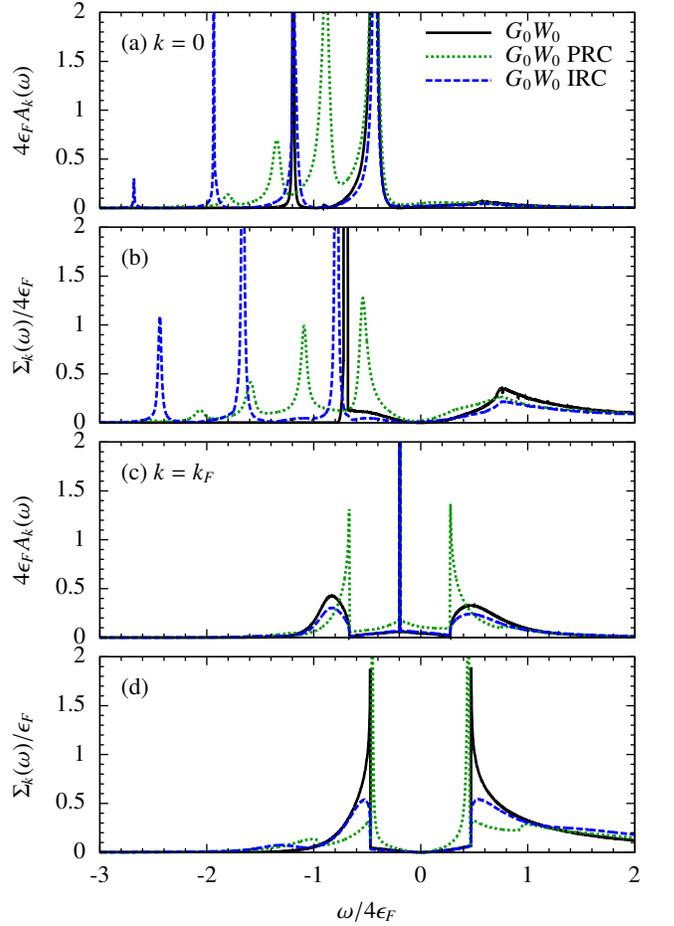}
\caption{A comparison of three distinct $GW$ schemes.
(a,c) The spectral weight for the 3D electron gas with 
$r_s = 4.0$. (b,d) The absolute value of the imaginary part of the 
proper self-energy, solved using Dyson's equation, corresponding 
to the spectral weights in (a) and (c).}
\label{fig:spectra}
\end{figure}

\begin{figure}[t!]
\centering
\includegraphics[scale = 1]{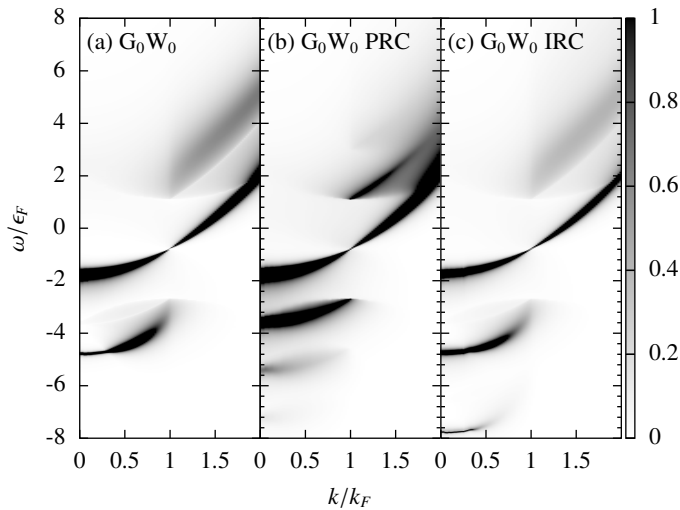}
\caption{Band structure for various $GW$ schemes
at $r_s = 4$. Note that the $G_0W_0$ IRC contains multiple satellites,
similar to the $G_0W_0$ PRC case, but with a spacing that mimics
the incorrect position of the $G_0W_0$ plasmaron peak.}
\label{fig:bands}
\end{figure}

\begin{figure}[b!]
\centering
\includegraphics[scale = 1]{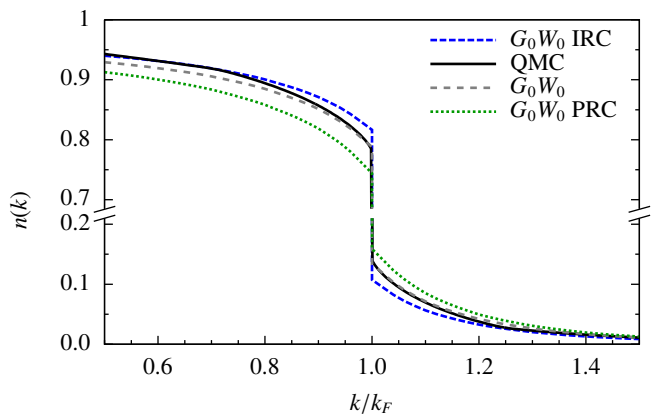}
\caption{The $k-$dependent occupation number $n(k)$ for all three 
approximation schemes at $r_s = 4.0$, plotted against the ``exact'' 
QMC result~\cite{Holzmann11,Olevano12}.}
\label{fig:momenta}
\end{figure}

\noindent where $|\rangle$ represents the direct product of the ground 
electronic state and all ground phonon states. The $G_0D_0$ IRC spectrum, 
although inaccurate in its peak spacing, still encodes the correct 
first energy moment. Thus, the IRC approach appears, within this simple 
toy model, to embody a compromise between the standard cumulant and 
self-energy $GW$ approaches. Furthermore, it appears not to corrupt some 
important aspects of the problem, such as the existence of multiple satellites 
and the value of the ``correlation'' energy.

The electron gas problem provides a more stringent and informative grounds of 
comparison for the different approximation schemes because no known 
method, including the $G_0W_0$ PRC scheme, is exact. Although the 
electron gas Hamiltonian differs significantly from the form of Eq.~\ref{phononh}, 
many of the observations made earlier about the three approximations 
to the Green's function should still apply. In Fig.~\ref{fig:spectra}a,c, we present 
the spectral weight 
\begin{equation}
A_k(\omega) = -{1 \over \pi}\t{Im }G_k^R(\omega)
\label{spectral_weight}
\end{equation}

\noindent for the 3D electron gas at density $r_s = 4.0$ both at the Fermi 
wavevector and at the bottom of the band. The spectral weight for the Dyson $GW$ self-energy 
is plotted for the same cases in Fig.~\ref{fig:spectra}b,d. The full band 
structure is reported for all three calculation schemes in Fig.~\ref{fig:bands}.

Both cumulant schemes improve the qualitative description of the satellite 
region at low $k$: whereas the $G_0W_0$ scheme predicts just one satellite 
at an energy $1.5\omega_p$ below the quasiparticle (QP) peak, both cumulant 
schemes predict a series of evenly-spaced satellites with decreasing weights. 
The $G_0W_0$ PRC scheme predicts that these satellite peaks are separated 
by approximately the plasmon energy $\omega_p$, whereas the $G_0W_0$ 
IRC scheme inherits the (presumably unphysical) $\sim$1.5$\omega_p$ spacing 
from the $G_0W_0$ calculation, exactly as it did in the electron-phonon model 
explored earlier. At and above the Fermi wavevector $k_F$, the $G_0W_0$ 
IRC results are much more similar to the $G_0W_0$ results
as compared to those from the $G_0W_0$ PRC. 
Overall, the $G_0W_0$ IRC scheme interpolates between the 
rounded spectral behavior of $G_0W_0$ at larger $k$ and the multiple satellite 
peak behavior of $G_0W_0$ PRC at lower $k$. 

In addition to the spacing of satellite peaks, another unphysical feature 
of the IRC approach is the appearance of spurious sharp quasiparticle resonances 
which are most apparent in the unoccupied portion of the spectral function. These 
features can be easily removed in a manner that hardly affects $A_k(\omega)$ for 
$k \leq k_F$, $n(k)$, or $\epsilon_{\t{corr}}/N$ (the latter two of which are discussed
next). The origin of these features and the means for their removal are discussed in 
the Appendix.

While the results for the 3D electron gas in the high frequency satellite wing 
suggest that the $G_0W_0$ PRC scheme is most accurate in this spectral region, 
these results shed little light on other properties, to which we now turn. We 
present in Fig.~\ref{fig:momenta} the wavevector-dependent occupation number
\begin{equation}
n_k = \int_{-\infty}^\mu A_k(\omega)d\omega.
\label{momenta}
\end{equation}

\noindent The $G_0W_0$ IRC scheme performs similarly to 
$G_0W_0$, erring on the opposite side of the ``exact'' quantum Monte 
Carlo (QMC) data~\cite{Holzmann11,Olevano12}. Notably, the $G_0W_0$ IRC occupation numbers 
match the QMC results almost exactly for $k < 0.8k_F$ and $k > 1.2k_F$; 
although the scheme suffers from unphysical satellite peak spacing at small 
$k$, it inherits rather accurate occupation numbers from $G_0W_0$. The 
$G_0W_0$ PRC occupation numbers are not as accurate, and yield a QP 
renormalization factor which is too small.

The accuracy of the occupation numbers in the $G_0W_0$ IRC 
approximation can be explained using the self-energy spectra in 
Fig.~\ref{fig:spectra}b,d. The most significant dependence of the 
momenta is on the weight of the main QP peak, which is determined 
by the slope of the real part of the self-energy. Since the real and 
imaginary parts of the self-energy are related by a Kramers-Kronig 
transform, the slope of the real part of the self-energy depends most 
strongly on the weight and positions of the peaks in the self-energy 
spectrum closest to $\omega \simeq \epsilon_k$. Since the $G_0W_0$ 
IRC self-energy spectrum peaks are more similar to the $G_0W_0$ ones, 
one would expect the $G_0W_0$ IRC momentum distribution near 
$k = k_F$ to more closely resemble that of $G_0W_0$. In particular, 
the increased QP renormalization factor with respect to the $G_0W_0$ 
approximation is explained by the slightly smaller peaks in the $G_0W_0$ 
IRC self-energy spectrum at $k = k_F$. 

\begin{table}[t!]
\begin{tabular*}{\linewidth}{@{\extracolsep{\fill}}  c  c  c  c  c  c }
  \hline
  \hline
  $r_s$ & $G_0W_0$ & $G_0W_0$ PRC & $G_0W_0$ IRC & QMC \\
  \hline   
  1 & $-0.0742$ & $-0.0688$ & $-0.0642$ & $-0.0600$\\
  2 & $-0.0542$ & $-0.0516$ & $-0.0467$ & $-0.0448$\\
  3 & $-0.0436$ & $-0.0411$ & $-0.0368$ & $-0.0369$\\
  4 & $-0.0375$ & $-0.0350$ & $-0.0310$ & $-0.0318$\\
  5 & $-0.0329$ & $-0.0304$ & $-0.0267$ & $-0.0281$\\
  \hline
  \hline
\end{tabular*}
\caption{Correlation energies of the 3D electron gas per particle 
$\epsilon_{\t{corr}}/N$ in Hartrees for various $GW$ 
schemes and $r_s$ values~\cite{Kas:17}. QMC values are
obtained from Vosco \emph{et al.}'s parameterization~\cite{Vosko80} of Ceperley
and Alder's fixed-node diffusion Monte Carlo data~\cite{Ceperley80}.}
\label{tab:g0w0}
\end{table}

\begin{figure}[t]
\centering
\includegraphics[scale = 1]{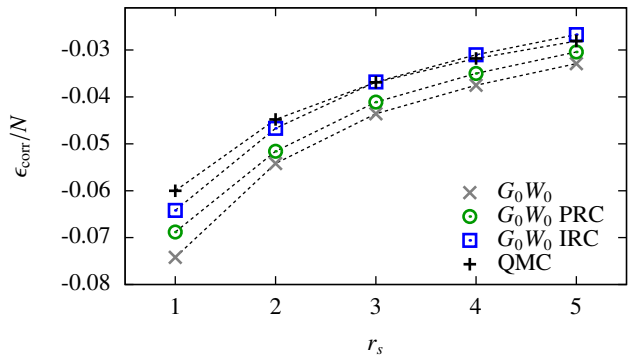}
\caption{Correlation energies of the 3D electron gas per particle $\epsilon_{\t{corr}}/N$
for various $GW$ schemes and $r_s$ values compared with the ``exact'' QMC result. Exact
values are reported in Table~\ref{tab:g0w0}.}
\label{fig:energies}
\end{figure}


Finally, total energies may be calculated from $A_k(\omega)$ using the Galitskii-Migdal formula
\begin{equation}
\epsilon = \sum_k\int_{-\infty}^\mu (\omega + \epsilon_k)A_k(\omega)d\omega,
\label{galitskii_migdal}
\end{equation}

\noindent where $\epsilon_k = k^2/2$ is the free-electron energy dispersion. 
The correlation energy per particle is calculated using 
$\epsilon_{\t{corr}} = (\epsilon - \epsilon_{\t{HF}})/N$, where 
$\epsilon_{\t{HF}}$ is the Hartree-Fock energy. For the cumulant 
schemes, $\mu$ is determined by enforcing the total particle number.
These energies are reported in Table I and plotted in Fig.~\ref{fig:energies}. 
For reference, the results from fixed-node diffusion Monte Carlo calculations
of Ceperley and Alder~\cite{Ceperley80}
are shown based on the parameterization by Vosko and coworkers~\cite{Vosko80}.
The $G_0W_0$ IRC scheme yields significantly more accurate 
correlation energies as compared to earlier schemes over this important range 
of $r_s$ values. 

\section{Conclusion}

In this work, we motivate the use of the improper retarded self energy 
in the cumulant generating function using Dyson's equation. Non-self-consistent 
calculations of the spectral weight show that the improper retarded cumulant 
(IRC) scheme predicts a series of multiple satellite bands with a 
larger-than-expected spacing at the bottom of the band. However, compared 
to the $G_0W_0$ PRC scheme, which predicts a series of satellite bands 
with a more physical $\omega_p$ spacing, the IRC scheme yields noticeably 
improved occupation numbers and correlation energies. 
This is promising in the ongoing research directed to unified, efficient
approaches for both total electronic energy and excitation energies.
Further work should be done to investigate other aspects related to the 
retarded cumulant-based $GW$ approaches discussed here, including 
self-consistency and the influence of higher-order cumulants.

\textit{Note added} -- Following  the submission of this work, several 
related studies have appeared~\cite{Gumhalter16,Caruso16,Vigil16}.
The spectral weights reported in these works for the electron gas model 
in the $G_0W_0$ PRC approximation are identical to the ones presented here.

\section*{Acknowledgements}
We thank J. Kas for numerous helpful discussions and aid with the implementation of numerics. M.Z.M. is supported by a fellowship from the National Science Foundation 
under grant number DGE-11-44155. Part of this work was done using 
resources from the Center for Functional Nanomaterials, which is a 
U.S. DOE Office of Science User Facility at Brookhaven National 
Laboratory under Contract No. DE-SC0012704 (MSH).

\section*{Appendix: Spurious Sharp Quasiparticle Resonances at $k > k_F$}

\begin{figure}[t]
\centering
\includegraphics[scale = 1]{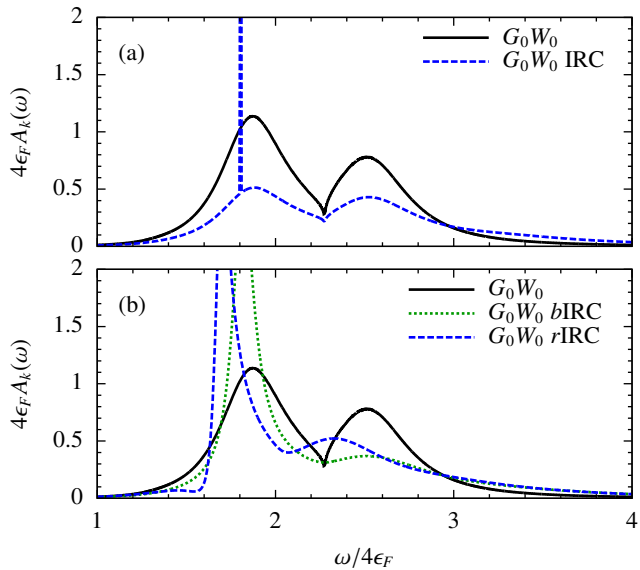}
\caption{Band structure for various $GW$ schemes
at $r_s = 4$ and $k = 3k_F$. The sharp quasiparticle resonance
predicted by $G_0W_0$ IRC is broadened by the $b$IRC and $r$IRC
schemes.}
\label{fig:highk_qp}
\end{figure}

The standard description of the cumulant function requires the evaluation of 
\begin{equation}
C_k(t) = {1 \over \pi}\int d\omega {|\t{Im} \Sigma(\omega + \epsilon_k)| \over \omega^2}(e^{-i\omega t} + i\omega t - 1),
\end{equation}

\noindent which for the IRC scheme demands that the integrand is constructed with $\t{Im }\Sigma^{I,R}(\omega + \epsilon_k)$. It should be noted, however, that the geometric sequence that is the improper self-energy is ill-defined when evaluated precisely on the energy shell since $G_k^0(\epsilon_k)$ is divergent there. This divergence implies $\t{Im }\Sigma^{I,R}(\epsilon_k) = 0$, resulting in a sharp quasiparticle-like feature in $A_k(\omega)$ superimposed on a smooth continuum. This feature is most apparent for $k > k_F$, as can be seen in Fig.~\ref{fig:highk_qp}a. Note that the smooth continuum behavior is much like that of the standard $G_0W_0$ spectral function.

This feature may be removed in a variety of ways that preserve all of the conclusions reached in the main text. For example, if the cumulant function $C_k(t)$ is defined such that the evaluation of $\t{Im }\Sigma_k\left(\omega + \epsilon_k^{G_0W_0}\right)$ is used as opposed to $\t{Im }\Sigma_k(\omega + \epsilon_k)$, then no spurious resonance appears (renormalized IRC or \emph{r}IRC). In addition, $A_k(\omega)$ for $k \leq k_F$, $n(k)$, and $\epsilon_{\t{corr}}/N$ are essentially unchanged. The same outcome occurs if the frequency of $G_k^0(\omega)$ within $\Sigma_k^{I,R}(\omega)$ is given an imaginary part equal to $\t{Im }\Sigma_k^*\left(\epsilon_k^{G_0W_0}\right)$, the quasiparticle lifetime associated with a standard $G_0W_0$ calculation (broadened IRC or \emph{b}IRC). 

\begin{table}[t]
\begin{tabular*}{\linewidth}{@{\extracolsep{\fill}}  c  c  c  c  c  c }
  \hline
  \hline
  $r_s$ & $G_0W_0$ IRC & $G_0W_0$ $b$IRC & $G_0W_0$ $r$IRC & QMC \\
  \hline   
  1 & $-0.0642$ & $-0.0643$ & & $-0.0600$\\
  2 & $-0.0467$ & $-0.0467$ & & $-0.0448$\\
  3 & $-0.0368$ & $-0.0363$ & & $-0.0369$\\
  4 & $-0.0310$ & $-0.0304$ & $-0.0308$ & $-0.0318$\\
  5 & $-0.0267$ & $-0.0260$ & & $-0.0281$\\
  \hline
  \hline
\end{tabular*}
\caption{Correlation energies of the 3D electron gas per particle 
$\epsilon_{\t{corr}}/N$ in Hartrees for the broadened IRC ($b$IRC)
and the renormalized IRC ($r$IRC) schemes.}
\label{tab:broad_energies}
\end{table}

These distinct regularization procedures do however alter the shape of $A_k(\omega)$ for $k > k_F$ as shown in Fig.~\ref{fig:highk_qp}b. Table~\ref{tab:broad_energies} shows the robustness of the correlation energies for the different schemes; we do not replot $A_k(\omega)$ for $k \leq k_F$ or $n(k)$ because they are essentially unaltered from Figs.~\ref{fig:spectra},\ref{fig:bands},\ref{fig:momenta}. Overall, we find that the most important features of the IRC approach are basically unmodified under the removal of these spurious resonances.


\begin{thebibliography}{47}
\bibitem [{Kas()}]{Kas:17}%
  \BibitemOpen
  \href@noop {} {\bibinfo  {journal} {Slight differences in the PRC energies
  from those presented in Ref.~\cite{Kas14} are due to differing underlying
  $GW$ self-energies. These differences are on the order of the numerical
  uncertainty in the correlation energy as calculated using
  Eq.~\ref{galitskii_migdal}}\ }\BibitemShut {NoStop}%
\end{thebibliography}
\end{document}